\documentclass{webofc}

\usepackage[varg]{txfonts}   
\usepackage{hyperref}
\usepackage{url}
\usepackage{caption}
\usepackage{subcaption}

\hypersetup{colorlinks=true,citecolor=blue,urlcolor=blue,linkcolor=blue}
\begin{document}
\title{Event-by-event vortex rings in fixed-target p+Ar collisions}
%
%

\author{\firstname{David} \lastname{Chinellato}\inst{1}\fnsep\thanks{speaker, \email{david.chinellato@gmail.com}}
\and \firstname{Michael} \lastname{Lisa}\inst{2}
\and \firstname{Willian} \lastname{Serenone}\inst{1}
\and \firstname{Chun} \lastname{Shen}\inst{3}\fnsep\thanks{\email{chunshen@wayene.edu}}
\and \firstname{Jun} \lastname{Takahashi}\inst{1}
\and \firstname{Giorgio} \lastname{Torrieri}\inst{1}
}

\institute{
Instituto de Fisica Gleb Wataghin, Universidade Estadual de Campinas, Campinas, Brasil
\and Department of Physics, The Ohio State University, Columbus, Ohio 43210, USA
\and Department of Physics and Astronomy, Wayne State University, Detroit, Michigan 48201, USA
}

\abstract{We present event-by-event simulations for central asymmetric p+Ar collisions at $\sqrt{s_\mathrm{NN}} = 68$ GeV to investigate the formation and evolution of vortex-ring structures from the early-stage longitudinal flow velocity profile. Our predictions for their imprints on Lambda hyperon's polarization observables are complementary to those presented in Ref.~[Phys.Rev.C 110 (2024) 5, 054908] and can be explored in the future fixed-target collisions at the Large Hadron Collider beauty (LHCb) experiment.
}
\maketitle
\section{Introduction}
\label{intro}

Understanding the role of large orbital angular momentum in non-central relativistic heavy-ion collisions has become an active frontier in high-energy nuclear physics~\cite{Becattini:2020ngo}. Beyond the measurements of global hyperon polarization at the Relativistic Heavy-Ion Collider (RHIC)~\cite{STAR:2017ckg}, the dynamics of local velocity gradient fields provide a rich picture of how the collision systems' angular momentum is deposited into the hydrodynamic medium, inducing local vorticity during the collective expansion, and how they manifest in the final-state observables~\cite{Betz:2007kg, Serenone:2021zef, Ribeiro:2023waz}. In asymmetric nuclear collisions — such as Cu+Au~\cite{Voloshin:2017kqp} or $p$+Au/Pb~\cite{Lisa:2021zkj} — the imbalance of nuclear thickness and stopping powers at different transverse positions of the light-ion projectile generates strong shear flows and longitudinal velocity gradients, which naturally seed a toroidal flow pattern (``smoke ring'') along the beam direction~\cite{Ivanov:2018eej}. It is intriguing to look for experimental observables that probe such topologically non-trivial flow structures. The production-plane polarization of $\Lambda$ hyperons, defined relative to the $\Lambda$'s momentum and the beam axis, is a promising observable for tracking the ``vortex-ring'' feature in asymmetric collision events. The vortex ring structure can be quantified with~\cite{Lisa:2021zkj, DobrigkeitChinellato:2024xph}
\begin{equation}
    \label{eq:RLambda3vectors}
   \mathcal{R}_{\Lambda}^{\hat{z}} \equiv
    2 \left\langle\frac{\vec{S}_{\Lambda} \cdot \left(\hat{z} \times \vec{p}_{\Lambda} \right)} {|\hat{z} \times \vec{p}_{\Lambda}|} \right\rangle_{\phi_\Lambda},
\end{equation}
where $\hat{z} \equiv (0, 0, 1)$ points in the direction of the light-ion beam, and the average is taken over the $\Lambda$ momentum azimuthal angle $\phi_\Lambda$ about the beam. The vectors $\vec{S}_{\Lambda}$ and $\vec{p}_{\Lambda}$ are the spin and momentum vectors for the $\Lambda$ hyperons in the lab frame, respectively.

From a phenomenological perspective, the effects of vortex rings are expected to be pronounced in heavy-ion collisions in the Beam Energy Scan program, where both asymmetric and lower-energy collisions are anticipated to amplify initial energy-momentum transport and vorticity formation~\cite{Ivanov:2018eej}. Studying vortex ring formation in asymmetric collisions also provides a novel phenomenological avenue to connect initial-state asymmetries with final-state polarization and flow observables. By comparing measurements across system size, beam energy, and centrality, it becomes possible to isolate the fingerprints of vortex rings and constrain the microscopic mechanisms responsible for generating and transporting vorticity in the QGP.

In this proceeding, we use a (3+1)D hydrodynamic model used in Ref.~\cite{DobrigkeitChinellato:2024xph} to provide complementary prediction for $\mathcal{R}_{\Lambda}^{\hat{z}}$ in p+Ar collisions at $\sqrt{s_\mathrm{NN}} = 68$ GeV, whose experiment can be performed at the LHCb System for Measuring Overlap With Gas (SMOG) setup.

\section{The (3+1)D dynamical framework}

We employ a geometric-based 3D initial condition developed in Refs.~\cite{Shen:2017fnn, Shen:2020jwv, Ryu:2021lnx, Alzhrani:2022dpi} connecting with a hydrodynamics + hadronic transport model. This initial-state model maps the local energy and longitudinal momentum from the colliding nucleons to hydrodynamic fields, ensuring the system's orbital angular momentum is conserved at the matching. The system's initial longitudinal flow can be parameterized with the nuclear thickness functions $T_{A(B)}(\vec{x}_\perp)$ as
\begin{equation}
    y_L(\vec{x}_\perp) = f y_\mathrm{CM}(\vec{x}_\perp) \quad\mbox{with}\quad y_\mathrm{CM}(\vec{x}_\perp) = \mathrm{arctanh}\left[\frac{T_A - T_B}{T_A + T_B} \tanh(y_\mathrm{beam}) \right],
    \label{eq:yL}
\end{equation}
where the parameter $f \in [0, 1]$ controls the portion of the initial net longitudinal momentum that is attributed to the flow velocity. Numerical simulations are carried out using the \texttt{iEBE-MUSIC} framework. The details of the hydrodynamic modeling are discussed in Ref.~\cite{DobrigkeitChinellato:2024xph}.
For spin-1/2 $\Lambda$ hyperons, their averaged spin vector over the freezeout hyper-surface $\Sigma_\mu$ can be computed as \cite{Becattini:2016gvu}
\begin{align}
    S^\mu(p^\mu) = - \frac{1}{8m} \frac{\int d^3 \Sigma_\alpha p^\alpha f_0(1 - f_0) \epsilon^{\mu\nu\alpha\gamma} p_\nu \omega_{\alpha \gamma}^\mathrm{th} }{\int d^3 \Sigma_\alpha p^\alpha f_0}.
    \label{eq:SpinVector}
\end{align}
Here, $f_0 \equiv (\exp[(E - \mu)/T] + 1)^{-1}$ is the Fermi-Dirac distribution for $\Lambda$ hyperons, $\omega_{\alpha \gamma}^\mathrm{th}$ is the thermal vorticity tensor, and $\epsilon^{\mu \rho \sigma \tau}$ is the Levi-Civita tensor. We choose the convention $\epsilon^{txyz} = 1$. We compute the spin vectors for $\Lambda$ and anti-$\Lambda$ hyperons. The difference between the two particle species lies in the opposite sign of the chemical potential in the thermal distribution function $f_0$. The net baryon chemical potential $\mu_B$ of individual fluid cells assigns different emission weights for $\Lambda$ vs. anti-$\Lambda$ in Eq.~\eqref{eq:SpinVector}, which result in different values in the final ring observables.
Because the ring observable in Eq.~\eqref{eq:RLambda3vectors} is driven by the transverse distribution of longitudinal velocity and density gradients, it is expected to be sensitive to the value of parameter $f$ in the initial-state model.

\section{Results}

\begin{figure}[h!]
    \centering
    \includegraphics[width=0.5\linewidth]{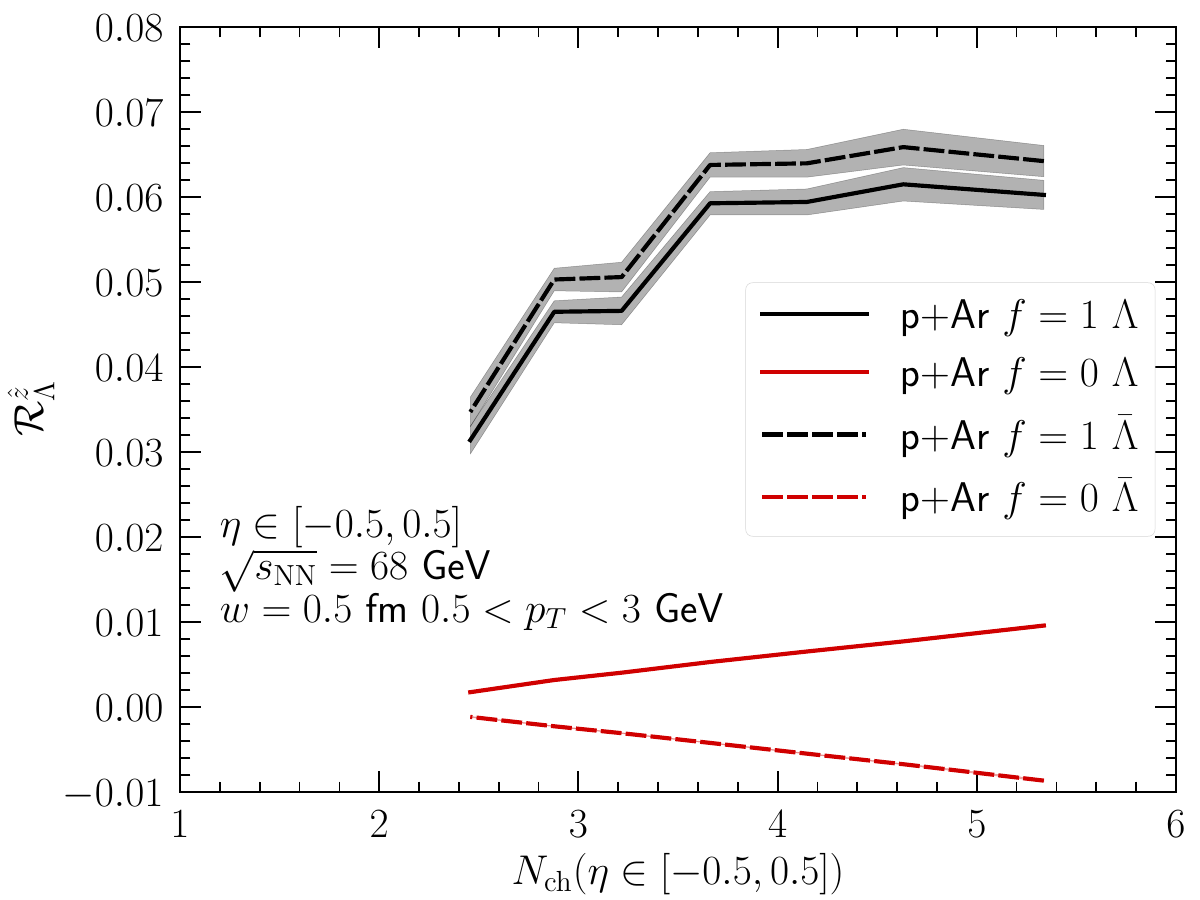}
    \caption{The vortex ring observable $\mathcal{R}_{\Lambda}^{\hat{z}}$ as function of charged hadron multiplicity for $\Lambda$ and anti-$\Lambda$ hyperons in p+Ar collisions at $\sqrt{s_\mathrm{NN}} = 68$ GeV with two different initial-state longitudinal flow profiles. The pseudo-rapidity is in the center-of-mass frame.}
    \label{fig:Rspin_Nchdep}
\end{figure}

\begin{figure}[h!]
    \centering
    \includegraphics[width=0.49\linewidth]{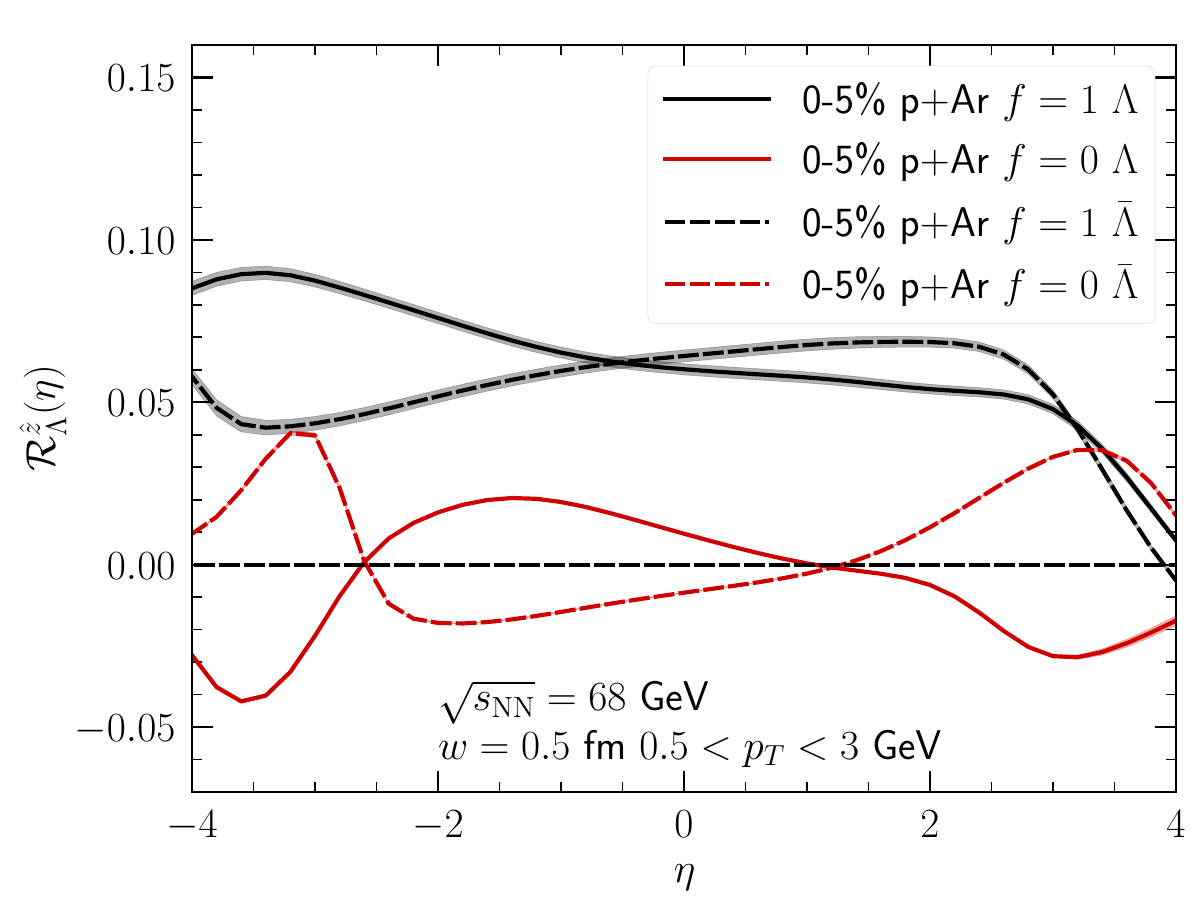}
    \includegraphics[width=0.49\linewidth]{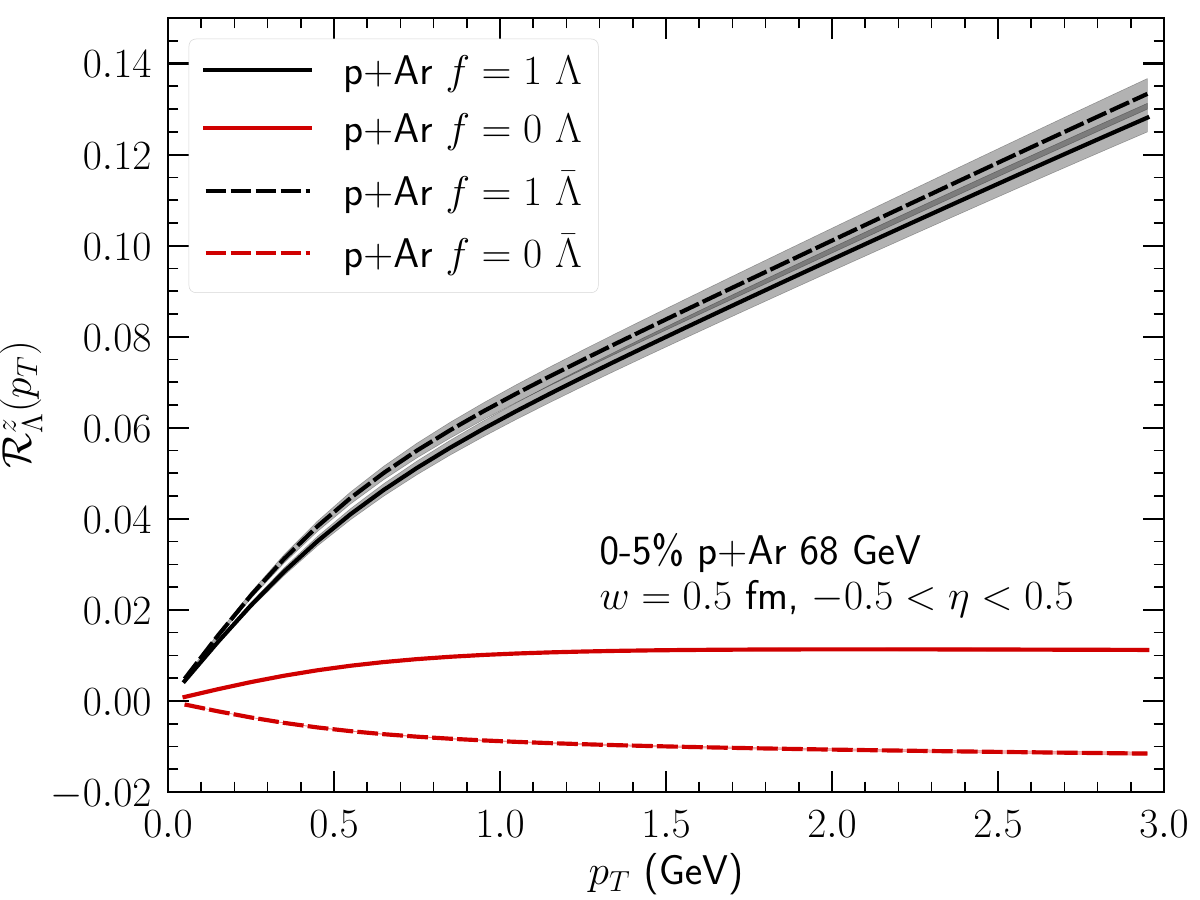}
    \caption{The vortext ring observables $\mathcal{R}_{\Lambda}^{\hat{z}}$ as functions of the center-of-mass pseudo-rapidity (left panel) and $p_T$ (right panel) for $\Lambda$ and anti-$\Lambda$ hyperons in p+Ar collisions at $\sqrt{s_\mathrm{NN}} = 68$ GeV.}
    \label{fig:Rspin_etapTdep}
\end{figure}

Figure~\ref{fig:Rspin_Nchdep} shows the integrated ring observable for $\Lambda$ and anti-$\Lambda$ near mid-rapidity as functions of the charged hadron multiplicity in p+Ar collisions. We observe that the $\mathcal{R}_{\Lambda}^{\hat{z}}$ is a factor of six larger with a strong initial longitudinal flow (the $f = 1$ scenario in Eq.~\eqref{eq:yL}) compared to the results with the Bjorken flow ($f = 0$). This result shows the strong sensitivity of the proposed ring observable to the early-time dynamics in the collision system. The magnitudes of $\mathcal{R}_{\Lambda}^{\hat{z}}$ increase with event activity (charged hadron multiplicity), demonstrating that the vortex ring flow structure develops with the produced fireball lifetime. Our simulations show that the ring observable for $\Lambda$ hyperons remains positive for both initial longitudinal flow cases. While the anti-$\Lambda$'s $\mathcal{R}_{\Lambda}^{\hat{z}}$ is negative with the zero initial longitudinal flow setup. To understand this difference, we need to study the differential dependencies of the ring observable as shown in Fig.~\ref{fig:Rspin_etapTdep}.
Figure~\ref{fig:Rspin_etapTdep} shows further details of the ring observables as functions of pseudo-rapidity in the center-of-mass frame and their transverse momentum. The $\mathcal{R}_{\Lambda}^{\hat{z}}(\eta)$ shows a rich structure of pseudo-rapidity dependence, demonstrating that the proposed ring observable could provide valuable insights into the collision system's longitudinal dynamics. The difference between $\Lambda$ and anti-$\Lambda$ originates from the $\mu_B$ distribution in the longitudinal direction. The $\mathcal{R}_{\Lambda}^{\hat{z}}(p_T)$ near mid-rapidity shows linear $p_T$ dependence for $p_T > 1$ GeV, whose slope is proportional to the thermal vorticity tensor~\cite{DobrigkeitChinellato:2024xph}.

\bigskip
\noindent {\it{Acknowledgments.}}
{
This work is in part supported by the U.S. Department of Energy (DOE) under award numbers DE-SC0021969 and DE-SC0020651.
C.S. acknowledges a DOE Office of Science Early Career Award.
M.L. acknowledges the support of the Fulbright Commission of Brazil.
J.T.~was supported by FAPESP projects 2023/13749-1 and CNPq through 303650/2025-7.
G.T.~acknowledges support from Bolsa de produtividade CNPQ 305731/2023-8, Bolsa de pesquisa FAPESP 2023/06278-2.
This research was done using resources provided by the Open Science Grid (OSG), which is supported by the National Science Foundation awards \#2030508 and \#1836650.
}

%
\bibliography{refs} 
%
%

\end{document}